\documentclass[aps,twocolumn,showpacs,superscriptaddress,reprint]{revtex4-2}

\usepackage{bm}
\usepackage{graphicx}
\usepackage{amsmath}
\usepackage{amsfonts}
\usepackage{siunitx}
\usepackage{bbold}
\usepackage{comment}
\usepackage{hyperref}

\renewcommand{\vec}[1] {{\bm{#1}}}
\newcommand{\ham}{\mathcal{H}}
\newcommand{\ce}[1] {$\mathrm{#1}$}

\newcommand{\up}{\uparrow}
\newcommand{\dw}{\downarrow}
\newcommand{\kp}{\vec{k}_\parallel}
\usepackage{mathtools}

\newcommand{\ket}[1] {\lvert #1{\rangle}}

\newcommand{\braket}[2] {{\langle #1{\lvert} #2{\rangle}}}

\usepackage{color}
\usepackage{colortbl}
\usepackage{array}
\usepackage{tikz}

\definecolor{cblue}{RGB}{ 138, 152, 255}%0,85,212}
\definecolor{cred}{RGB}{  255, 168, 138 }

\begin{document}

\title[Effect of the external fields in high Chern number quantum anomalous Hall insulators]%
{Effect of the external fields in high Chern number quantum anomalous Hall insulators}

\author{Yuriko Baba}

\affiliation{GISC, Departamento de F\'{\i}sica de Materiales, Universidad Complutense, E--28040 Madrid, Spain}

\author{Mario Amado}

\affiliation{Nanotechnology Group, USAL--Nanolab, Universidad de Salamanca, E--37008 Salamanca, Spain}

\author{Enrique Diez}

\affiliation{Nanotechnology Group, USAL--Nanolab, Universidad de Salamanca, E--37008 Salamanca, Spain}

\author{Francisco Dom\'{\i}nguez-Adame}

\affiliation{GISC, Departamento de F\'{\i}sica de Materiales, Universidad Complutense, E--28040 Madrid, Spain}

\author{Rafael A. Molina}

\affiliation{Instituto de Estructura de la Materia, IEM-CSIC, E--28006 Madrid, Spain}


\begin{abstract}

A quantum anomalous Hall state with high Chern number has so far been realized in multiplayer structures consisting of alternating magnetic and undoped topological insulator layers. However, in previous proposals, the Chern number can be only tuned by varying the doping concentration or the width of the magnetic topological insulator layers. This drawback largely restrict the applications of dissipationless chiral edge currents in electronics since the number of conducting channels remains fixed. In this work, we propose a way of varying the Chern number at will in these multilayered structures by means of an external electric field applied along the stacking direction. In the presence of an electric field in the stacking direction, the inverted bands of the unbiased structure coalesce and hybridize, generating new inverted bands and collapsing the previously inverted ones. In this way, the number of Chern states can be tuned externally in the sample, without the need of modifying the number and width of the layers or the doping level. We showed that this effect can be uncovered by the variation of the transverse conductance as a function of the electric field at constant injection energy at the Fermi level.

\end{abstract}

\maketitle

\section{Introduction}
%- Review QAH: https://arxiv.org/abs/2205.15226

%- Analytical model inspired by~\cite{Wang2013} but applied to the case of multilayer heterostructues:~\cite{Wang2021}

%- Thin slabs + electric field: theory (\cite{Duong2015, Wang2015}) experiment~\cite{Zhang2017a}

%- Finite size effect: theory~\cite{Fu2014}, experiment~\cite{Feng2016}

Topological materials show great promise for future applications in electronics. The various members of the family of topological materials manifest a wide range of interesting properties, but the most significant one is the dissipationless electric transport through chiral edge or surface states. The introduction of the resistance standard based on the quantum Hall effect is a salient example of the applications of topological properties of materials as it was a groundbreaking advance in metrology~\cite{vonKlitzing2017}.   

The family of quantum Hall effects now includes the quantum spin Hall effect~\cite{Maciejko2011} and the quantum anomalous Hall effect~\cite{Chi2022}. These effects do not require an external magnetic field to occur and the main difference between them is the presence or absence of time-reversal symmetry. While classical anomalous Hall effect was discovered in ferromagnetic materials by Hall himself in the 1880s~\cite{Hall1880}, its quantum counterpart was the last of the family of quantum Hall effects to be measured and this happened only recently. The first experiments used magnetically doped topological insulator thin films~\cite{Chang2013QAH}, where the 
the magnetic impurities provided the breaking of time-reversal symmetry needed to observe the anomalous Hall effect. 

A new development of the quantum anomalous Hall effect has been achieved in heterostructures made of topological insulator layers with and without magnetic doping. Depending on the number of layers, the Chern number $C$, the topological quantity measuring the number of edge states, can be controlled at will~\cite{Wang2013,Fang2014}. Chern numbers as high as $C=5$ have been measured in experiments using this technique~\cite{Zhao2020}. 

On the other hand, external electromagnetic fields can be used to manipulate the topological properties of materials. In fact, dc electric fields can induce a topological phase transition depending on their strength in topological insulators~\cite{Kim2012,Zhang2013a} and as well as in topological semimetals~\cite{Pan2015,Collins2018}. Application of external electric fields can renormalize the Fermi velocity and control the decay of surface states in Dirac materials\cite{Diaz-Fernandez2017,Diaz-Fernandez2017b,Baba2019}. Time-dependent electromagnetic fields can also be used for controlling topological properties of matter~\cite{Lindner2011,Cayssol2013,Usaj2014,Dehghani2015,Chan2016,Gonzalez2016,Gonzalez2017,McIver2020}. These studies have also been extended to magnetic topological insulators, showing clear signatures of the quantum anomalous Hall effect~\cite{Duong2015,Wang2015,Zhang2017a}.

In the case of high-Chern number magnetic insulators, it would certainly be desirable to tune the topological properties in the same sample by means of electric fields. In fact, an external electric field could be of use in different types of electronic devices where a fast change in the amount of current or the effective electrical resistance are needed. The goal of this work is to study the effect of an electric field in the topological properties of heterostructures with magnetic doping showing the anomalous Hall effect with high-Chern numbers. In particular, we will focus on the experimentally feasible heterostructures made of Bi$_2$Se$_3$ with layers with Cr doping. We will present Chern number maps as a function of the heterostructure parameters and different values of the electric field. These maps display a rich structure, allowing us to fine tuning the values of the Chern number, the number of edge states and, consequently, the transport properties of the device. We have also studied the evolution of the spatial localization of the electron states along the stacking and transverse directions as a function of the electrical field to have a deeper understanding of how Chern number changes take place.

\section{Model} \label{sec:model}

The system under consideration consists of alternated magnetically doped and undoped layers of the same topological insulator, as depicted in Fig.~\ref{fig:setup}. In order to study the electron states in the topological insulator \ce{Bi_2Se_3}, we use the low energy Hamiltonian introduced in Ref.~\cite{Liu2010}. We write it in the basis of the bands of bonding and antibonding $p_z$ orbitals ordered as $\ket{P_1^+\up}$, $\ket{P_2^+\dw}$, $\ket{P_1^+\dw}$, $\ket{P_2^-\up}$, where the superscripts $\pm$  stand for even and odd parity, and $\up, \dw$ represent spin-up and spin-down states, respectively. In this basis, the Hamiltonian reads
\begin{equation} \label{eq:mod:ham}
\ham (z, \kp) = 
    \left[\begin{matrix}
    H_+(z, \kp) & B k_z \sigma_y \\
    B k_z \sigma_y & H_-(z, \kp) 
    \end{matrix}\right]~,
\end{equation}
where $H_\pm (\vec{k})$ describes two blocks of $2\times 2$ models with opposite chirality, hybridized by the terms proportional to $B$, which are linear in $k_z$. The $H_\pm$ are given by the following expression
\begin{align}  \label{eq:mod:hampm}
    H_\pm(z, \kp ) = & ~ \epsilon(\kp, z) \mathbb{1}_2  + \left[M (z, \kp) \mp g(z) \right] \tau_3  \nonumber
    \\& + A k_x \tau_1 \pm A k_y \tau_2~, 
\end{align}
with $g(z)$ being the Zeeman splitting at position $z$ along the stacking direction and $\mathbb{1}_n$ the $n\times n$ unit matrix. In Eqs.~\eqref{eq:mod:ham} and~\eqref{eq:mod:hampm}, $\sigma_i$ and $\tau_i$ denote Pauli matrices acting in the spin basis and the basis of $P_{1}^{+}$ and $P_{2}^{-}$ subbands, respectively. The on-site term and the mass term are respectively given by $\epsilon(z, \kp) = C_{0} - C_{1} \partial_{z}^{2} + C_{2} \left(k_{x}^{2} + k_{y}^{2}\right)$ and $M(z, \vec{k})= M_0(z) - M_{1} \partial_{z}^{2} + M_{2} \left(k_{x}^{2} + k_{y}^{2}\right)$. The topology of the Hamiltonian is given by the sign of the mass parameters: if $M_0<0$ and $M_1, M_2>0$, the system is in the inverted regime. On the other hand, the diagonal term $\epsilon(z, \kp)$ accounts for the particle-hole asymmetry and has no impact in the topological nature of the bands. Therefore, $\epsilon(z, \kp)$ can be set to zero without loss of generality.

Due to the \ce{Cr^-} magnetic doping, a Zeeman splitting is induced in the magnetic region~\cite{Yu2010} and the band inversion tends to reduce~\cite{Zhang2013}. Therefore, the mass $M_0(z)$ and the Zeeman splitting $g(z)$ parameters of the heterostructure are modelled by step-like functions along the stacking direction, following the same dependence as in Ref.~\cite{Zhao2020}. In the undoped region $g(z)=0$ and the mass term takes the pristine value $M_0(z) = M_0$. In the \ce{Cr^-} doped region, $M_0(z) = M_0^\mathrm{Cr}$ and $g(z) = g$, where both values are tuned by the doping concentration. Notice that $g>0$ and, in Eq.~\eqref{eq:mod:ham}, we set the splitting to be equal in both orbitals but opposite sign depending on the spin.

Finally, an external electric field is added as a linear potential in the stacking direction given by
\begin{equation} \label{eq:hamf}
    \ham_\mathrm{f} = e f z \mathbb{1}_4~,
\end{equation}
where $e$ is the elementary electric charge and $f$ is the external electric field. 

\section{Results}

\begin{figure}[htb]
\begin{comment}
\begin{tabular}{ m{0.5cm} m{5cm} }% *{2}{>{\centerin\arraybackslash}b{\dimexpr0.5\linewidth-2\tabcolsep\relax}}}
\begin{tikzpicture}
  \node (A)              at (0,-1) {$z$};
  \draw[-stealth]        (0,-0.7)   -- (0,0);
\end{tikzpicture} & 
\renewcommand{\arraystretch}{1.3}
 \begin{tabular}{m{0.8\linewidth}}
 \cellcolor{cblue} ~Cr-doped TI $30~\mathrm{\AA}$\\ 
 \cellcolor{cred} ~undoped TI $40~\mathrm{\AA}$ \\
 \cellcolor{cblue} ~Cr-doped TI $30~\mathrm{\AA}$\\ 
 \cellcolor{cred} ~undoped TI $40~\mathrm{\AA}$ \\
 \cellcolor{cblue} ~Cr-doped TI $30~\mathrm{\AA}$\\ 
\end{tabular}
\end{tabular}
\end{comment}
\includegraphics[width=0.5\columnwidth]{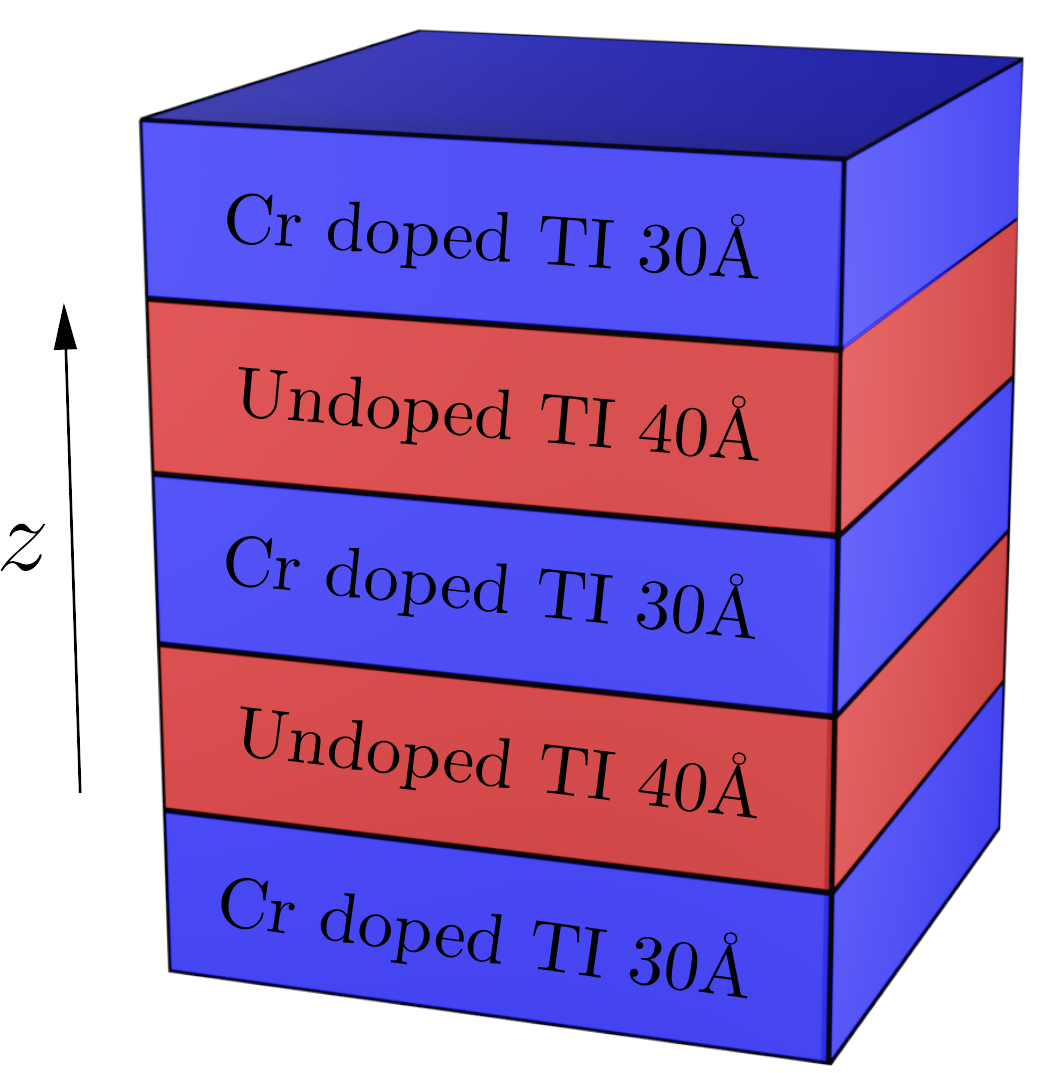}
\caption{Schematic representation along the stacking direction $z$ of alternated Cr-doped ($30\,$\AA) and undoped films ($40\,$\AA) of the topological insulator (TI) heterostructure.}
\label{fig:setup}
\end{figure}

For concreteness, we focus on the heterostructure reported in Ref.~\cite{Zhao2020} and shown schematically in Fig.~\ref{fig:setup}. The layered structure under consideration consists of three films of three quintuple layers (3QL) of \ce{(Bi,Sb)_{2-x}Cr_x Te_3} with two films of four quintuple layers (4QL) of \ce{(Bi,Sb)_2 Te_3} in between. In the following, we will refer to the Cr-doped 3QL layers as Cr-QL and to the pristine topological insulator 4QL as TI-QL, whose widths are \SI{30}{\angstrom} and \SI{40}{\angstrom}, respectively.

\begin{figure}[ht]
    \centering
    \includegraphics[width=0.9\textwidth]{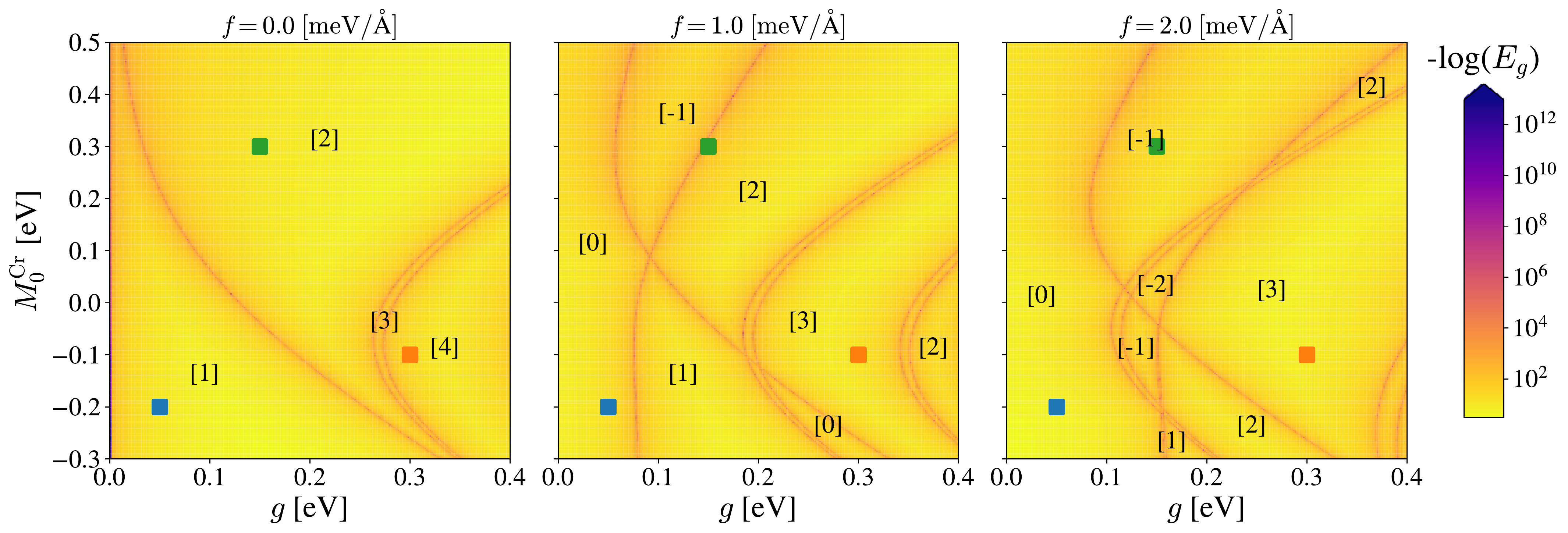}
    \caption{Phase diagram of the energy gap, in logarithmic scale, and the Chen numbers of the system indicated between the regions of gap closing. Panels (a), (b) and (c) correspond to electric fields of $\SI{0}{\milli\electronvolt/\angstrom}$, $\SI{1}{\milli\electronvolt/\angstrom}$ and $\SI{2}{\milli\electronvolt/\angstrom}$, respectively. The blue, orange and green squares correspond to the calculated points in Fig.~\ref{fig:Cond}.}
    \label{fig:Map1D}
    % ##TODO: add letters to panels. 
\end{figure}

This configuration is highly symmetric and enables the access to high Chern numbers states up to 4 in the absence of electric field. As shown in Figure \ref{fig:Map1D}(a), the phase diagram of the system comprises wide regions where $C = 1$, $2$ and $4$ and a small region with $C=3$. Notice that the region with $C=4$ is inconsistent with the results presented in Ref.~\cite{Zhao2020}, as already pointed out by Wang \emph{et al.}~\cite{Wang2021}. However, we find that a small region with $C=3$ is actually present in between those with $C=2$ and $C=4$, in contrast to Ref.~\cite{Wang2021}. The inversion symmetry in the $z$ direction assures that the states come in pairs, except in the case of the fully interlayer-connected $C=1$ state; due to finite size effect and the subsequent discretization of levels, the $C=3$ region pops out in tight-binding models. 

Notice that the phase diagrams are plotted as a function of both $g$ and $M_0^{\mathrm{Cr}}$ as independent parameters. However, doping concentration is expected to modify both quantities~\cite{Yu2010, Zhang2013} in a correlated way. Therefore, the complete phase diagram is not accessible experimentally as long as multiple samples must be grown in order to modify the doping concentration and the Chern number. Due to this issue, an external field that changes dynamically the properties of a given device is a remarkable alternative to tuning the Chern number. More precisely, an electric field in the direction of growth does not break the translation symmetry of the system in $x$ and $y$ directions, enabling the Chern number to be computed in an effective 2D quantum Hall insulator. In Fig.~\ref{fig:Map1D}, we plot the evolution of the gap phase map with the electric field. We use the method implemented in \texttt{Z2Pack}~\cite{Gresch2017, Soluyanov2011}, based in hybrid Wannier functions, to compute the Chern number numerically in the gapped regions for a half-filled spectrum, i.e. $E_F = 0$ in our model. The gap is computed by diagonalization of the tight binding Hamiltonian for a 1D chain with $z$ stacking shown in Fig.~\ref{fig:setup} and zero in-plane momentum.

The electric field beaks the translation symmetry along the $z$ direction but it does not break the particle-hole symmetry, as long as it has the same impact on all the orbitals. Therefore, the spectrum is still particle-hole symmetric and the degenerate branches diverge when they acquire a non-zero expectation value $\langle{z}\rangle$. In fact, the electric field acts as a confining potential on the states due to the Stark effect and tends to close the gap by the coalescence of particle and hole bands with opposite spatial distribution. A simple picture of the band behaviour with the electric field can be obtained from perturbation theory. If we consider the term $\ham_\mathrm{f}$ of Eq.~\eqref{eq:hamf} as a perturbation of the Hamiltonian given by Eq.~\eqref{eq:mod:ham}, the first order correction in the energy is
\begin{equation} \label{eq:pert}
    \delta E^{(1)} = \braket{\psi^0_{n, \vec{k}}}{\ham_\mathrm{f}| \psi^0_{n, \vec{k}} }~,
\end{equation}
where $\psi^0_{n, \vec{k}}$ is the unperturbed state of the $n$th Bloch band. Considering the linear potential given by Eq.~\eqref{eq:hamf}, the first order correction is an energy shift proportional to the expectation value $z$ and the electric field
\begin{equation}
    \delta E^{(1)} = e f \braket{\psi^0_{n, \vec{k}}}{z| \psi^0_{n, \vec{k}} }~.
\end{equation}
Therefore, the states with $\langle{z}\rangle$ positive (negative) increase (decrease) their energy. Inside the gap, this implies the coalescence of the bands of holes and electrons with opposite sign of $\langle{z}\rangle$. The bulk states remain almost unchanged, while the energy shift affects specially the states peaked in the interfaces between layers due to their localization. 

\begin{figure}[ht]
    \centering
    \includegraphics[width=0.9\textwidth]{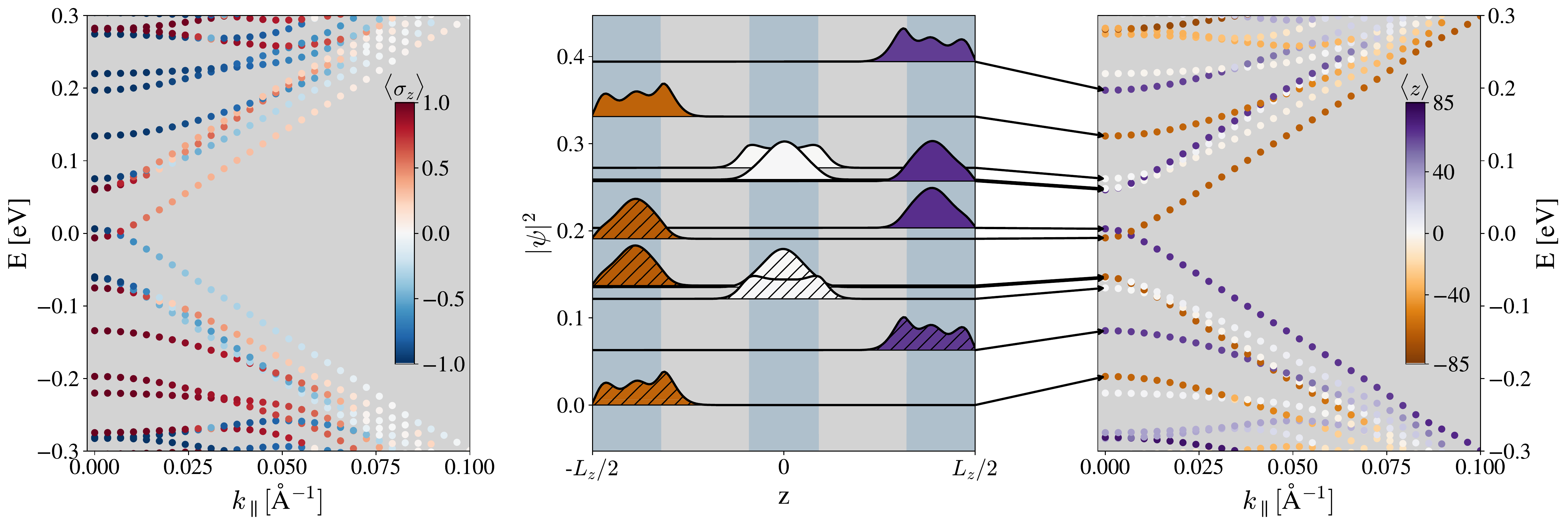}
    \caption{Energy dispersion and states for a $C=3$ state at $f = \SI{1}{\milli\electronvolt/\angstrom}$ in a 1D chain in the $z$ direction. The parameters are $M_0^\mathrm{Cr} = \SI{-0.2}{\electronvolt}$ and $g=\SI{0.32}{\electronvolt}$. The energy dispersions show the projection of $\sigma_z$ and $\langle z \rangle$ in (a) and (c) panels, respectively. Panel (b) shows the probability density $|\psi| ^2$ for the indicated states at $k_\parallel=0$. The states below the Fermi energy are hatched. The parameters correspond to a $C=4$ state in the absence of electric field (see Fig.~\ref{fig:States0mV} for comparison with the case at $f=0$). }
    \label{fig:States5mV}
\end{figure}

The topological nature of the bands of this multilayered systems is encoded in the band inversion phenomenon occurred between pairs of the conduction and valence bands with different spin polarizations~\cite{Wang2013}. In fact, even if counterintuitive, the spatial localization of the wave functions along the $z$ direction is not a definite signature of the topology of the bands~\cite{Wang2021} (see the Appendix \ref{App:localization} for more details). In Fig.~\ref{fig:States5mV} the states of a $C=4$ system are plotted in the presence of an electric field of $\SI{1}{\milli\electronvolt/\angstrom}$. By comparing with the case for zero electric field reported in Fig.~\ref{fig:States0mV}, it can be noticed that the states that are peaked at the sides of the slide break their inversion symmetry and localize in one of the sides, lifting also the degeneracy between levels. Therefore, within the addition of an electric field, even if the Chern number is still given by the crossing between levels, the localization plays a role because it gives the direction and strength of the displacement of the bands. 

\onecolumngrid

\twocolumngrid

\onecolumngrid

\twocolumngrid

\onecolumngrid

\begin{figure}[ht]
    \centering
    \includegraphics[width=0.9\textwidth]{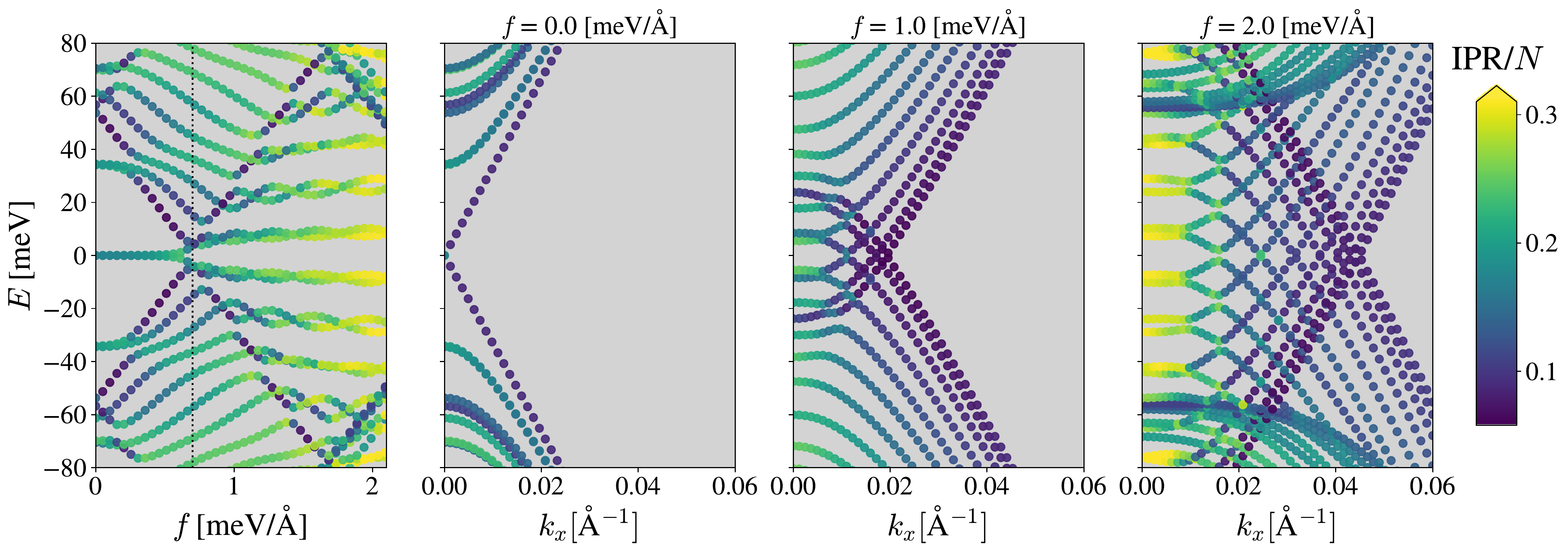}
    \caption{Energy levels at zero momentum as a function of the electric field (a) and energy dispersion for $f =  0, 1, 2 \si{\meV/\angstrom}$, in panels (b), (c) and (d) in a 2D slab with $L_y = \SI{400}{\angstrom}$ and $z$ stacking corresponding to Fig.~\ref{fig:setup}. The parameters are $M_0^\mathrm{Cr} = \SI{-0.2}{\electronvolt}$ and $g=\SI{0.05}{\electronvolt}$, corresponding to a $C=1$ in the absence of electric field, changing to trivial at $f\sim \SI{0.7}{\electronvolt / \angstrom}$, the $f = \SI{0.7}{\electronvolt / \angstrom}$ is indicated in a dashed vertical line in panel (a). The dispersions also show the $\mathrm{PR}/N$ as an indicator of the localization of the state. The $\mathrm{PR}$ is calculated with Eq.~\eqref{eq:IPR}.}
    \label{fig:IPRC1}
     % ## TODO: check labels of the axis
\end{figure}

\twocolumngrid

\onecolumngrid

\begin{figure}[ht]
    \centering
    \includegraphics[width=0.9\textwidth]{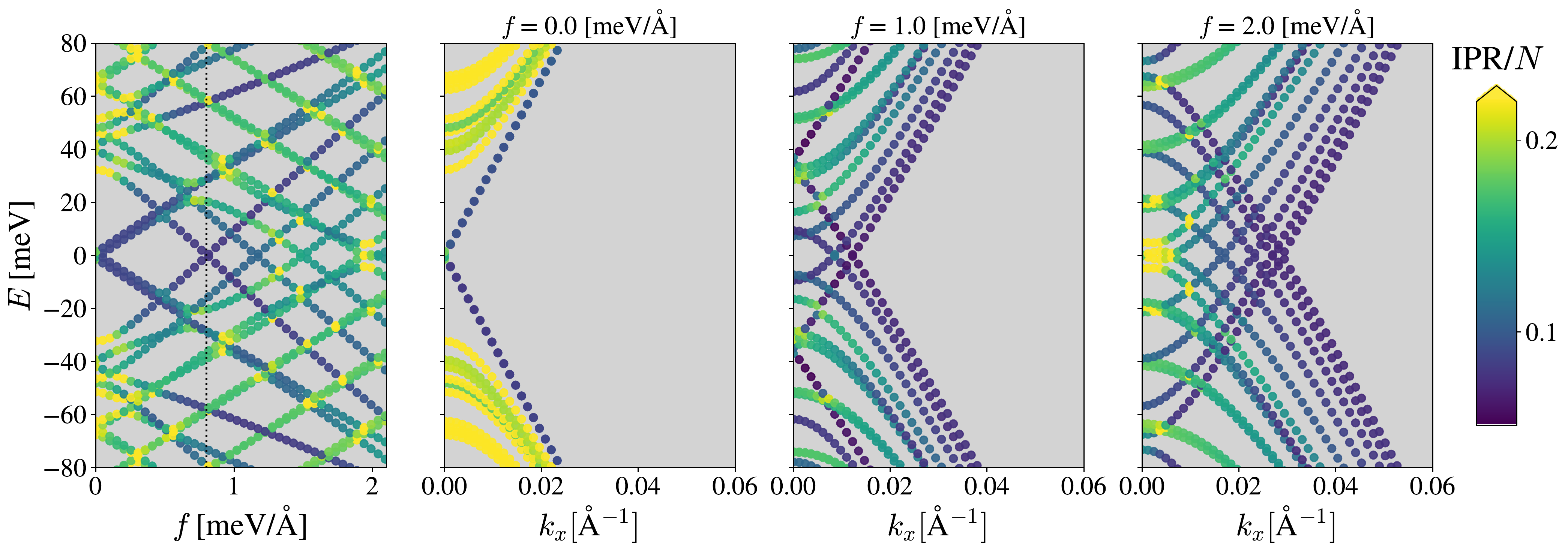}
    \caption{(a)~Energy levels at zero momentum as a function of the electric field and energy dispersion for (b)~$f =\SI{0}{\milli\electronvolt/\angstrom}$, (c)~$f =\SI{1}{\milli\electronvolt/\angstrom}$ and (d)~$f =\SI{2}{\milli\electronvolt/\angstrom}$ in a 2D slab with $L_y = \SI{400}{\angstrom}$ and $z$ stacking corresponding to Fig.~\ref{fig:setup}. The PR projection is plotted in the scale indicated on the right side of the plots. 
    The parameters are $M_0^\mathrm{Cr} = \SI{0.3}{\electronvolt}$ and $g=\SI{0.1}{\electronvolt}$, corresponding to a $C=2$ state that shifts to $C=-1$ at $f\sim \SI{ 0.8 }{\electronvolt / \angstrom}$. The value of $f = \SI{ 0.8}{ \electronvolt / \angstrom} $ is indicated in a dashed vertical line in panel (a).}
    \label{fig:IPRC2}
\end{figure}

\twocolumngrid

Even if the states are modified by the electric field, the main reason for the Chern number tuning is the energy shift given by Eq.~\eqref{eq:pert} and not the spatial localization induced by the field. In fact, the effect of the localization in a single slab of \ce{Cr^-}doped TI is not very relevant as well as the mass renormalization given by the localization (see Appendix \ref{App:localization} and \ref{App:slab} for more details). However, if we consider a slab of finite area in transverse direction, the topological states appear inside the gap and they can be identified by their localization in the $x-y$ plane. In Fig.~\ref{fig:IPRC1} and~\ref{fig:IPRC2} we show two representative cases of the evolution of the energy as a function of the electric field. In the case plotted in Fig.~\ref{fig:IPRC1}, a $C=1$ state is turned trivial due to the hybridization of the bands at $f \sim \SI{0.7}{\electronvolt/\angstrom}$, while in Fig.~ \ref{fig:IPRC2}, the case of a $C=2 $ going to $C=-1$ is obtained by both the hybridization of the topological bands and by the crossing of a new level coming from the upper part of the energy spectrum. 

In the previously mentioned figures, the bands are plotted with the projection of the participation ratio~(PR) in the basis of the spatial coordinates, defined as follows
\begin{equation} \label{eq:IPR}
    \mathrm{PR} = \frac{\left(\sum_{i=1}^{N} p_i\right)^2}{\sum_{i=1}^{N} p_i^2}~,
\end{equation}
where $N$ is the number of sites in the system and $p_i$ is the probability density at site $i$, i.e. $|\psi_i|^2$. With this definition, the maximum values of $\mathrm{PR}$ is $N$ for a completely delocalized state and $1$ for a state localized in only one site. Although the spatial localization is not a sufficient condition for establishing the topological character of electron states, these states are by their own nature localized in the edges of the sample. Therefore, the PR value is an indicative quantity in order to track the evolution of the states with the electric field, specially in finite-size slabs where the states evolve in a continuous way with the electric field, in contrast with the calculated abrupt transitions in the semi-infinite Chern number results represented in figure~\ref{fig:Map1D}. 
 
\begin{figure}
    \centering
    \includegraphics[width=0.7\linewidth]{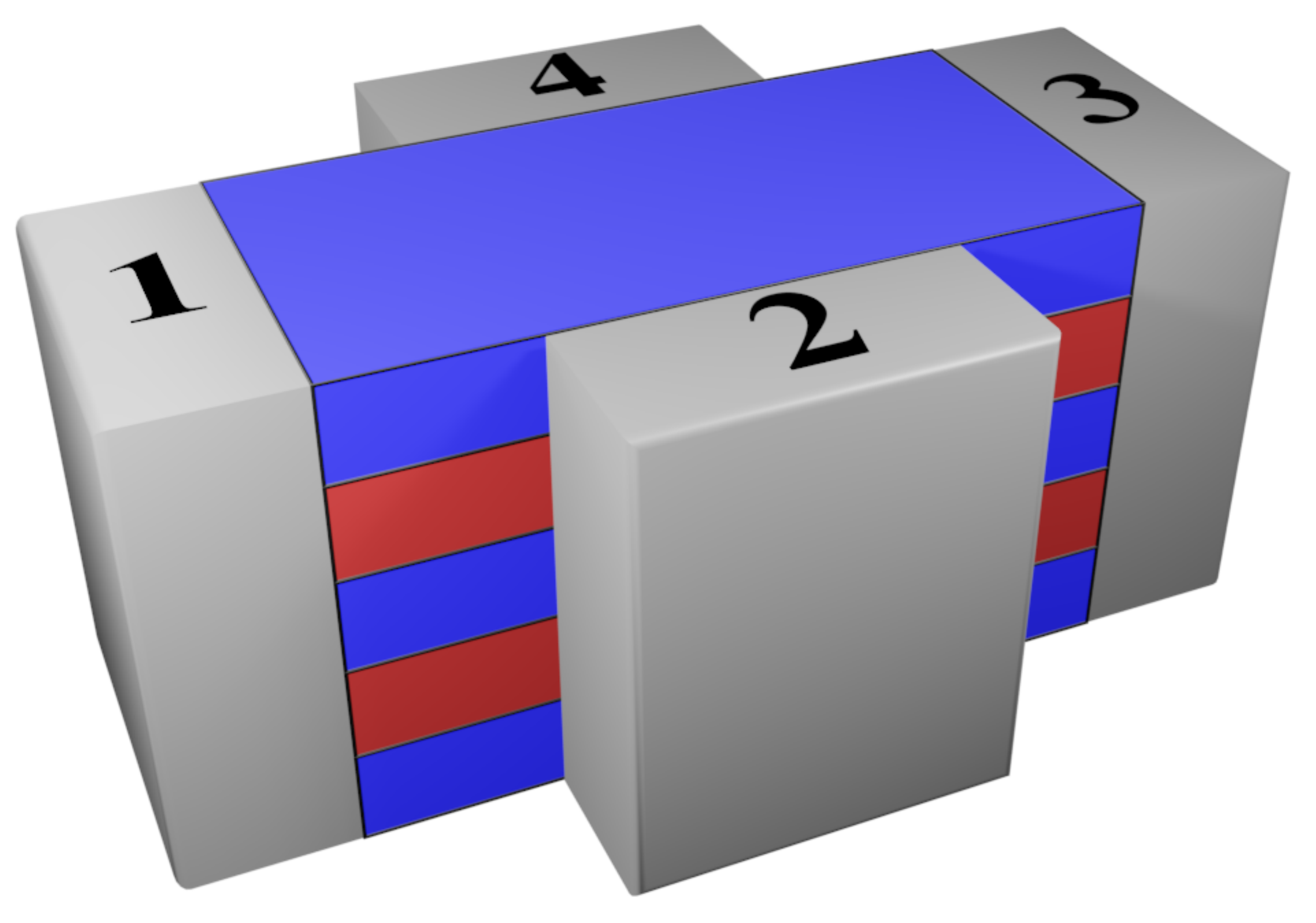}
    \caption{Sketch of the Hall bar for the conductance simulations. The size is $L_x =\SI{300}{\angstrom}$ and $L_y = \SI{180}{\angstrom}$. In the stacking direction ($z$), the sample resembles the same geometry shown in Fig.~\ref{fig:setup}. } 
    \label{fig:CondDevice}
\end{figure}

Finally, we address the problem of measuring these topological phase transitions in a transport set-up. An experimentally feasible scenario is a minimal Hall bar with four terminals, where the occurrence of transverse conductance is a key signature of the topological states. Figure~\ref{fig:CondDevice} shows a schematic view of the setup. The system comprises the multilayered sample connected to four leads with the same stacking of layers. The electric field is applied only in the scattering region, excluding the leads, and an Anderson-like disorder is added in order to take into account defects and other imperfections of the sample. The Anderson disorder is introduced in the Hamiltonian by adding the following term
\begin{equation} \label{eq:hamAnderson}
    \ham_\mathrm{A} = w (x,y,z)~\mathbb{1}_4~,
\end{equation}
where $w(x,y,z)$ is a function that at each site gives the onsite random energy, uniformly distributed in the range $[-W/2, W/2]$, with the disorder strength $W$. The transport simulations have been performed using the package \texttt{Kwant}~\cite{Groth2014} within the Landauer-B\"{u}ttiker formalism in a slab of $300 \times 180 \times 170\,\si{\angstrom}$. The injection energy is set to the Fermi energy with a small positive shift in order to exceed the finite-size gap. 

\begin{figure}
    \centering
    \includegraphics[width=0.8\linewidth]{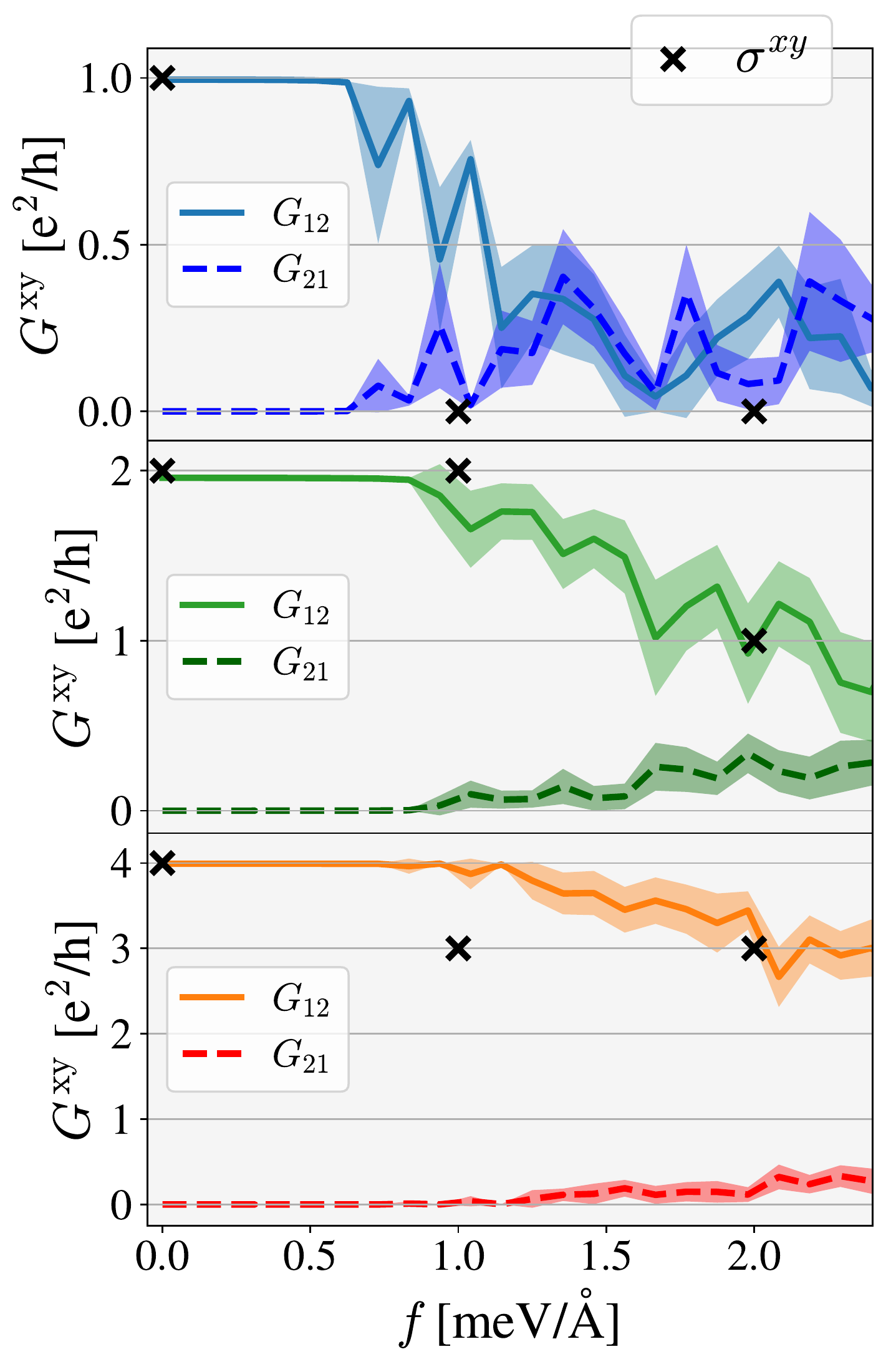}
    \caption{Conductance as a function of the electric field in a Hall bar with four terminals. Lines indicate the average transverse conductance while shadowed areas correspond to the squared variance for $100$ disorder realizations.Solid lines indicate the conductance $G_{12}$ while dotted lines are the conductance in reversed direction $G_{21}$. Small crosses are the expected values of the transverse conductance given by the Chern number from Eq.~\eqref{eq:HallChern}. The parameters considered correspond to the coloured squares in the phase diagram plotted in Fig.~\ref{fig:Map1D}, given by the following set of parameters: $g=\SI{0.05}{\electronvolt}$ and $M_0^\mathrm{Cr}=\SI{-0.2}{\electronvolt}$ (blue), $g=\SI{0.15}{\electronvolt}$ and $M_0^\mathrm{Cr}=\SI{0.3}{\electronvolt}$ (green), $g=\SI{0.32}{\electronvolt}$ and $M_0^\mathrm{Cr}=\SI{-0.2}{\electronvolt}$ (orange).}
    \label{fig:Cond}
\end{figure}

Figure \ref{fig:Cond} shows the conductance for three different sets of parameter as a function of the electric field. The states that contribute to the transverse conductance are mostly the Chern states. Additionally, the hybridized bulk-surface states can contribute to the transverse transport as well. Even so, disorder has a significant impact on these trivial states and their conductance is mostly reduced upon increasing the disorder strength. In the aforementioned figure, the average conductance is plotted together with the squared variance of the disorder realizations for an Anderson disorder with strength $W = 4 E_g$, where $E_g$ is the energy of the first mode above the Chern states in the absence of electric field and is a measure of the effective finite-size gap. 

If the transverse conductance is mediated only by Chern states, it must fulfil the following relation
\begin{equation} \label{eq:HallChern}
    G_\mathrm{Chern} = \frac{e^2}{h} \sum_{n} C (n)~,
\end{equation}
where the sum is computed over the Chern numbers $C(n)$ of the occupied bands only, i.e. the states below the Fermi energy. In Fig.~\ref{fig:Cond}, the value of the expected Hall conductance from Eq.~\eqref{eq:HallChern} is indicated by the crosses. Notice that the average conductance exceeds the expected quantized values due to the contribution of the bulk states that appears at the Fermi energy when the electric field closes the gap. In fact, before the first gap closing, the conductance shows a perfect quantization even in the presence of disorder. At the first gap closing, the conductance starts to decrease smoothly and the disorder affects the transmission of the trivial states, yielding higher values of the variance.

Due to the breaking of the time reversal symmetry with the addition of the Zeeman splitting, the current of the anomalous Hall channels flows in a preferential direction. If we use the numbering of leads introduced in Fig.~\ref{fig:CondDevice} and denote the conductance direction by the order of the subscript indices such that $G_{pq} = G_{p\rightarrow q}$, we get that $G_{12}$ is the preferential current flow direction, while $G_{21}$ is mediated by the contributions of trivial currents. 

The measure of the transmission in the opposite direction gives us a criterion to establish the robustness of the topological states with the electric field as well. In Fig.~\ref{fig:Cond}, the current in the direct direction is indicated in solid line, while the reversed current is indicated in dotted line. Notice that, even if the current in the opposite direction becomes non-zero with the electric field, its value is always reduced in comparison with the $G_{12}$ if the there are topological states. In the blue panel in Fig.~\ref{fig:Cond} $G_{21}$ and $G_ {12}$ have the same order of magnitude due to the triviality of the phase.

% direction of the current is indicated by the order of the 
% Quantized currenmt from 0 to 2 (0 out)

\section{Conclusions}

We have studied the effect of an external electric field on the states of a heterostructure of magnetically doped topological insulator. The multiple subbands generated by the stacking of alternating Cr-doped and undoped layers enable the occurrence of several inverted bands and, by means of this mechanism, multiple Chern states can be achieved. In the presence of an electric field in the stacking direction, the bands coalesce and hybridize, generating new inverted bands and collapsing the previously inverted ones. In this way, the number of Chern states can be tuned externally in the sample, without the need of modifying the number and width of the layers or the doping level. However, due to the multiple band crossing possibilities, the topological transitions are difficult to predict and the phase diagram obtained are complex and intricate. The level of complexity is increased by finite-size effects in a slab, where the hybridization of edge states located at opposite surfaces is enhanced due to the finite width of the sample. 

In all this phenomenology, the general trend is to decrease the number of inverted bands by the collapsing of the topological states. We showed that this effect can be measured in the variation of the transverse conductance as a function of the electric field at constant injection energy at the Fermi level. In our simulations, defects and other imperfections of the device are taken into account by an Anderson-like model of disorder, which enables a better visualization of the topological features by decreasing the contribution of the trivial bulk states. 

% In conclusion, we presented a convenient way to induce a topological phase transition in high Chern number QAH states by an external field. 

\appendix

\section{Spatial localization and effective models}  \label{App:localization}

The layers stacked in the $z$ direction exhibit topological states that are peaked typically at the interfaces between them. However, as proved by Wang and Li~\cite{Wang2021}, the spatial distribution is not evidence of the topology of the bands, which instead is determined by the inversion of their gaps. As already mentioned in the section \ref{sec:model}, in the paper by Wang \textit{et al.}~\cite{Wang2013}, a method of inspection of the effective mass have been proposed and successfully applied to a \ce{Cr^-}doped slab by decoupling the two chiral sectors and Fourier transform the Hamiltonian. More precisely, the idea is to set the coupling term $B=0$ and write the Hamiltonian in the basis of $\varphi_n(z) = \sqrt{2/L_z} \sin \left( n \pi z/L_z + n \pi/2\right)$, $n$ baing an integer. This way, one gets $N$ effective subbands with quantized $k_z = n \pi/ L_z$ and effective mass given by
\begin{align}  \label{eq:a1:mass}
    m_{\pm}(n) & =  \braket{\varphi_n}{M_0(z)|\varphi_n} + M_1 (n \pi/L_z)^2 \\
    & + M_2 (k_x^2+ k_y^2) \mp \braket{\varphi_n}{g(z)|\varphi_n} ~.  \nonumber
\end{align}

In the case of a single slab of \ce{Cr^-}doped TI, the $M_0(z) = M_0^\mathrm{Cr}$ and $g(z) = g$ and both magnitudes have no dependence on the spatial coordinate. Therefore, the effective mass reduces to a function that depends only on the model parameters and the quantized momentum in $z$: $m_\pm (n) = M_0 + M_1 (n \pi/L_z)^2 + M_2 (k_x^2+ k_y^2) \mp g$. The bands contribute with Chern number $+1$ if $m_+(n)<0$, $-1$ if $m_-(n)<0$ and they are trivial otherwise. The Chern number can be computed by summing up all the contributions by the following expression
\begin{equation}
    C = N_+ - N_-~,
\end{equation}
where $N_\pm$ is the number of bands with $m_\pm(n)<0$. This approximation is valid only for the case of $B=0$. However, when realistic parameters are added, the phase diagram obtained by this method must be adiabatically modified. 

In the case of multilayered system, the procedure is equivalent but the expected values in Eq.~\eqref{eq:a1:mass} depends on the distribution of the Cr-doped layers encoded in the spatial dependence of $M_0(z) $ and $g(z)$. Moreover, due to the dependence on the $z$ coordinate, the mass terms can generate non-diagonal terms in the Hamiltonian in correspondence to different subbands. Therefore, once the mass terms~\eqref{eq:a1:mass} are computed, the phase diagram must be adiabatically fitted by the non-diagonal terms and later modified by the term $B$ (see reference~\cite{Wang2021} for more details). 

The complexity of the heterostructure makes the approximation of the effective model more complex, but still feasible. The electric field generates similar non-diagonal terms and the effective mode approximation cannot be applied easily. Then, the Chern number must be computed numerically, by tools such as \texttt{Z2Pack}~\cite{Gresch2017}.

\begin{figure}
    \centering
    \includegraphics[width=0.7\linewidth]{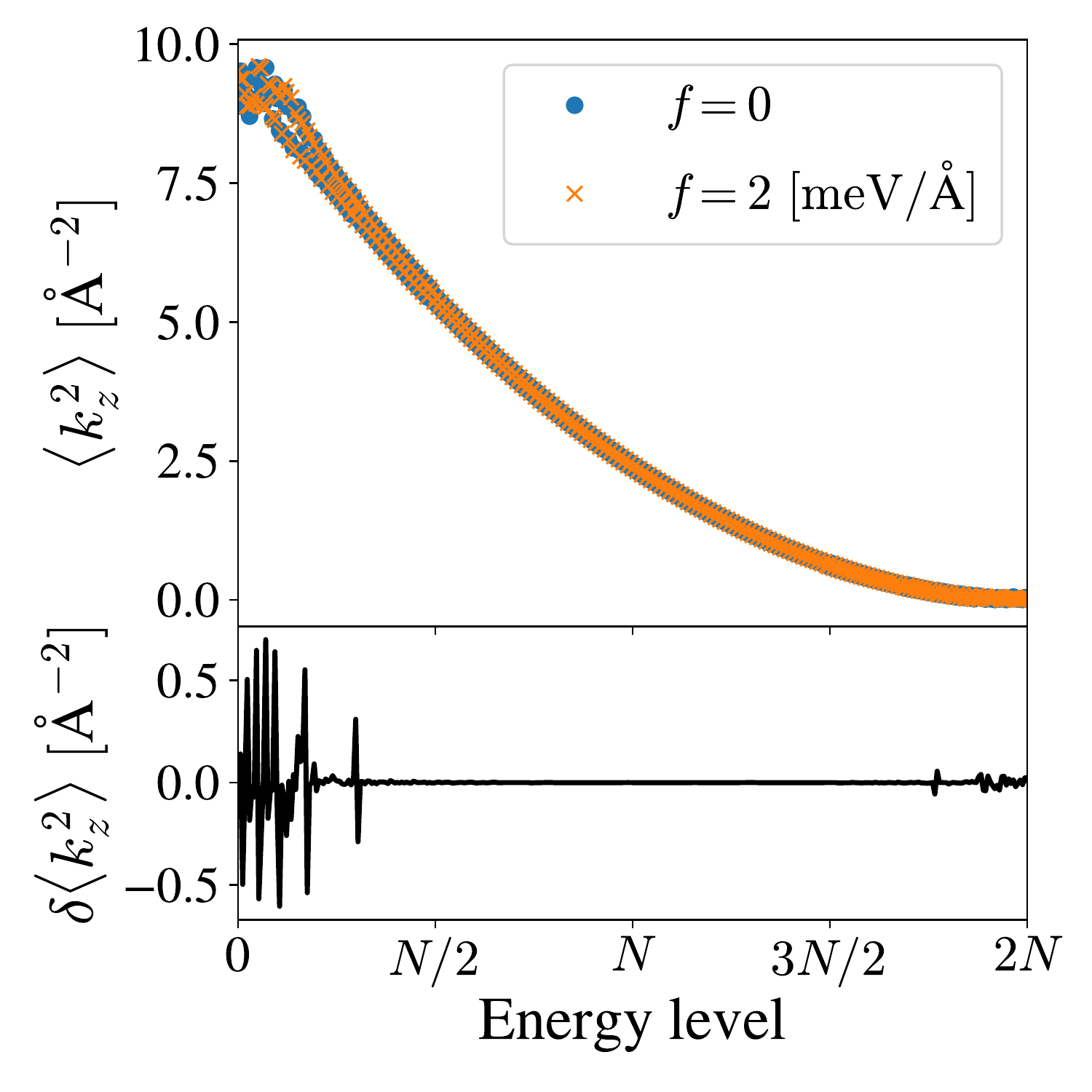}
    \caption{Expected value of $k_z^2$ for a multilayered system with parameters $M_0^\mathrm{Cr} = \SI{-0.2}{\electronvolt}$ and $ g = \SI{0.32}{\electronvolt}$.}
    \label{fig:effectivekz2}
\end{figure}

In Fig.~\ref{fig:effectivekz2}, the expected value of $k_z^2$ is plotted for the $2N$ bands below the Fermi level in a multilayered system with parameters corresponding to a $C=4$ state in the absence of electric field. Notice that the value of $k_z^2$ decreases by approaching the gap, but it is not modified by the external electric field. The lower panel shows the difference between the expected values in the presence of the electric field and in the pristine case for each band, proving that the modification of $\langle k_z^2 \rangle$ is almost negligible. In conclusion, the change of the effective mass in Eq.~\eqref{eq:a1:mass} by the varying of momentum does not play a significant role in the  modification of the Chern number by the external field. 

% m-n = 2j-1 siendo j entero

\onecolumngrid\
\begin{figure}
    \centering
    \includegraphics[width=\textwidth]{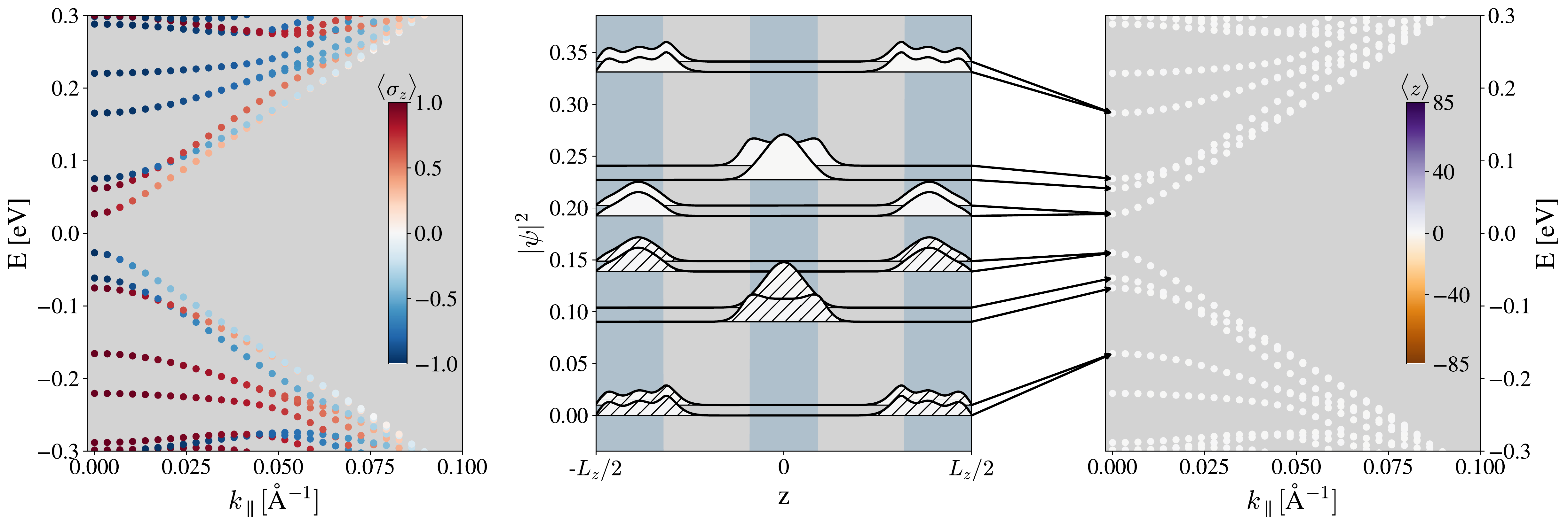}
    \caption{Energy dispersion and states for a $C=4$ state in the absence of electric field. The system is finite only in the $z$ direction and the parameters are $M_0^\mathrm{Cr} = \SI{-0.2}{\electronvolt}$ and $g=\SI{0.32}{\electronvolt}$. The energy dispersions show the projection of $\sigma_z$ and $\langle z \rangle$ in (a) and (c) panels, respectively. Notice that, due to the symmetry of the structure, the expected value of $z$ is zero for all the states. In panel (b), the probability density $|\psi|^2$ is plotted for the indicated states at $k_\parallel=0$. The states below the Fermi energy are hatched.}
    \label{fig:States0mV}
\end{figure}
\twocolumngrid\

\section{Effect of the electric field in a Cr-doped slab} \label{App:slab}

\begin{figure}
    \centering
    \includegraphics[width=\linewidth]{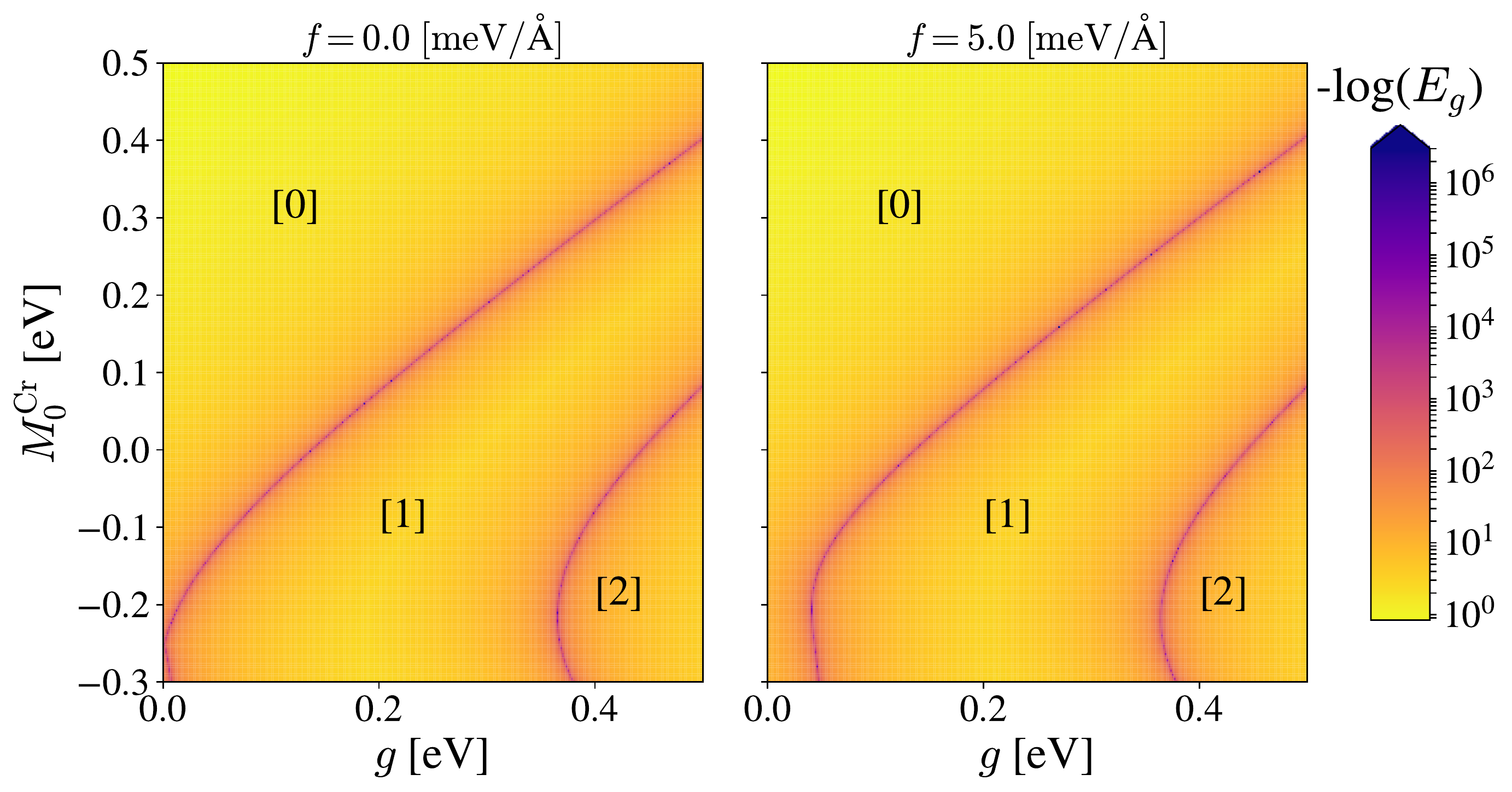}
    \caption{Phase diagram for a slab of width $\SI{30}{\angstrom}$ as a function of the mass parameter $M_0^\mathrm{Cr}$ and the Zeeman splitting $g$.}
    \label{fig:slabPhaseMap}
\end{figure}

In these section, the effect of the electric field on the effective mass is further explored by discussing the case of a Cr-doped slab. The modification of the bands by both the electric field and the Cr$^{-}$ doping in a slab of \ce{Bi_2Te_3} and \ce{Sb_2Te_3} have been already studied in the literature~\cite{Duong2015, Wang2015, Zhang2017a}. 
For comparison with the heterostructure results, we present the phase diagram for a slab of width $\SI{30}{\angstrom}$ and constant $M_0^ \mathrm{Cr}$ in Fig.~\ref{fig:slabPhaseMap}. Notice that, due to the strong confinement in the $z$ direction, the electric field only shifts slightly the regions of Chern number in the phase diagram without substantial modifications. As long as the heterostructure is obtained by the stacking of layers of similar size, we can conclude that the modification of the phase diagram with the applied electric field in Fig.~\ref{fig:Map1D} is mostly induced by the crossing of the subbands rather than from the change of the effective Hamiltonian parameters by the electric field. 

\acknowledgments

This work was supported by Spanish Ministry of Science and Innovation under grant PID2019-106820RB-C21/22 and by grant PGC2018-
094180-B-I00 funded by MCIN/AEI/10.13039/501100011033 and FEDER "A way of making Europe".

\end{document}